\documentstyle[10pt,epsf,epsfig,dp_delphititle,oldlfont,lineno,amssymb]{dp_delphi}
\makeindex
\def\DpPaperGroup{EP}
\def\DpPaperRef{2003-015}
\def\DpDate{1 April 2003}
\def\DpAuthors{DELPHI Collaboration}
\def\DpSubmit{(Accepted by Phys.Lett.B)}
\def\DpTitle{{ Study of Inclusive \jpsi\ Production \\
in Two-Photon Collisions at LEP~II \\
with the DELPHI Detector}}
\def\DpComment{ }
\def\DpEMail{ }
\def\jpsi{\hbox{$J{\kern-0.24em}/{\kern-0.14em}\psi$}}
\def\ev#1#2{\hbox{#1e{\kern-0.10em}V{\kern-0.30em}/{\kern-0.14em}$#2$}}
\def\gevc{\ev{G}{c}}
\def\gevcc{\ev{G}{c^2}}
\def\mevc{\ev{M}{c}}
\def\mevcc{\ev{M}{c^2}}
\def\eqn#1{(\ref{#1})}

\catcode`@=11
\def\eqalign#1{\null\,\vcenter{\openup\jot\m@th
\ialign{\strut\hfil$\displaystyle{##}$&$\displaystyle{{}##}$\hfil
\crcr#1\crcr}}\,}
\catcode`@=12
\def\bqt#1#2\eqt{\begin{equation}\label{#1}%
{#2}\end{equation}\noindent}
\def\bln#1#2\eln{\begin{equation}\label{#1}%
\eqalign{#2}\end{equation}\noindent}
\def\brlist{\begin{thebibliography}{99}}
\def\brf{\bibitem}
\def\erlist{\end{thebibliography}}
\def\bra{\langle}
\def\ket{\rangle}
\begin{document}
\makeatletter
\newcount\@tempcntc
\def\@citex[#1]#2{\if@filesw\immediate\write\@auxout{\string\citation{#2}}\fi
  \@tempcnta\z@\@tempcntb\m@ne\def\@citea{}\@cite{\@for\@citeb:=#2\do
    {\@ifundefined
       {b@\@citeb}{\@citeo\@tempcntb\m@ne\@citea\def\@citea{,}{\bf ?}\@warning
       {Citation `\@citeb' on page \thepage \space undefined}}%
    {\setbox\z@\hbox{\global\@tempcntc0\csname b@\@citeb\endcsname\relax}%
     \ifnum\@tempcntc=\z@ \@citeo\@tempcntb\m@ne
       \@citea\def\@citea{,}\hbox{\csname b@\@citeb\endcsname}%
     \else
      \advance\@tempcntb\@ne
      \ifnum\@tempcntb=\@tempcntc
      \else\advance\@tempcntb\m@ne\@citeo
      \@tempcnta\@tempcntc\@tempcntb\@tempcntc\fi\fi}}\@citeo}{#1}}
\def\@citeo{\ifnum\@tempcnta>\@tempcntb\else\@citea\def\@citea{,}%
  \ifnum\@tempcnta=\@tempcntb\the\@tempcnta\else
   {\advance\@tempcnta\@ne\ifnum\@tempcnta=\@tempcntb \else \def\@citea{--}\fi
    \advance\@tempcnta\m@ne\the\@tempcnta\@citea\the\@tempcntb}\fi\fi}
 
\makeatother
\begin{titlepage}
\pagenumbering{roman}
\CERNpreprint{\DpPaperGroup}{\DpPaperRef} 
\date{{\small\DpDate}} 
\title{\DpTitle} 
\address{\DpAuthors} 
\begin{shortabs} 
\noindent
Inclusive \jpsi\  production
in photon-photon collisions has been
observed at LEP~II beam energies.
A clear signal
from the reaction $\gamma\gamma\to\jpsi+X$ is seen.
The number of observed $N(\jpsi\rightarrow \mu^+\mu^-)$ events 
is $36\pm7$ for an integrated luminosity of 
617 pb$^{-1}$\kern-0.9em,\kern+0.9em yielding a cross-section
of $\sigma(\jpsi+X)=45\pm9{\rm (stat)}\pm17{\rm (syst)}\ {\rm pb.}$
Based on a study of the event shapes of different types of $\gamma\gamma$
processes in the PYTHIA program, we conclude that
$(74\pm22)$\% of the observed \jpsi\ events are due to 
`resolved' photons, the dominant contribution of which is most probably
due to the gluon content of the photon.

\end{shortabs}
\vfill
\begin{center}
\DpSubmit \ \\ 
\DpComment \ \\
\DpEMail \ \\
\end{center}
\vfill
\clearpage
\headsep 10.0pt
\addtolength{\textheight}{10mm}
\addtolength{\footskip}{-5mm}
\begingroup
%
\catcode`@=11
\def\eqalign#1{\null\,\vcenter{\openup\jot\m@th
\ialign{\strut\hfil$\displaystyle{##}$&$\displaystyle{{}##}$\hfil
\crcr#1\crcr}}\,}
\long\def\@makefntext#1{
\vskip0pt\parindent0pt\begin{list}{}%
{\labelwidth1.5em\leftmargin=\labelwidth%
\labelsep3pt\itemsep0pt\parsep0pt\topsep-2pt
\def\baselinestretch{1.0}\footnotesize}%
\item[\hfill\@makefnmark]#1\end{list}}
\long\def\@makecaption#1#2{\vskip10pt
{#1:\ \begingroup\small\baselineskip14pt plus4pt minus2pt
#2\par\endgroup}}
\catcode`@=12
\def\bln#1#2\eln{\begin{equation}\label{#1}%
\eqalign{#2}\end{equation}\noindent}
\newcommand{\DpName}[2]{\hbox{#1$^{\ref{#2}}$},\hfill}
\newcommand{\DpNameTwo}[3]{\hbox{#1$^{\ref{#2},\ref{#3}}$},\hfill}
\newcommand{\DpNameThree}[4]{\hbox{#1$^{\ref{#2},\ref{#3},\ref{#4}}$},\hfill}
\newskip\Bigfill \Bigfill = 0pt plus 1000fill
\newcommand{\DpNameLast}[2]{\hbox{#1$^{\ref{#2}}$}\hspace{\Bigfill}}
%
\footnotesize
\noindent
\DpName{J.Abdallah}{LPNHE}
\DpName{P.Abreu}{LIP}
\DpName{W.Adam}{VIENNA}
\DpName{P.Adzic}{DEMOKRITOS}
\DpName{T.Albrecht}{KARLSRUHE}
\DpName{T.Alderweireld}{AIM}
\DpName{R.Alemany-Fernandez}{CERN}
\DpName{T.Allmendinger}{KARLSRUHE}
\DpName{P.P.Allport}{LIVERPOOL}
\DpName{S.Almehed}{LUND}
% ...
\normalsize
\endgroup
\titlefoot{Department of Physics and Astronomy, Iowa State
     University, Ames IA 50011-3160, USA
    \label{AMES}}
\titlefoot{Physics Department, Universiteit Antwerpen,
     Universiteitsplein 1, B-2610 Antwerpen, Belgium \\
     \indent~~and IIHE, ULB-VUB,
     Pleinlaan 2, B-1050 Brussels, Belgium \\
     \indent~~and Facult\'e des Sciences,
     Univ. de l'Etat Mons, Av. Maistriau 19, B-7000 Mons, Belgium
    \label{AIM}}
\titlefoot{Physics Laboratory, University of Athens, Solonos Str.
     104, GR-10680 Athens, Greece
    \label{ATHENS}}
\titlefoot{Department of Physics, University of Bergen,
     All\'egaten 55, NO-5007 Bergen, Norway
    \label{BERGEN}}
\titlefoot{Dipartimento di Fisica, Universit\`a di Bologna and INFN,
     Via Irnerio 46, IT-40126 Bologna, Italy
    \label{BOLOGNA}}
\titlefoot{Centro Brasileiro de Pesquisas F\'{\i}sicas, rua Xavier Sigaud 150,
     BR-22290 Rio de Janeiro, Brazil \\
     \indent~~and Depto. de F\'{\i}sica, Pont. Univ. Cat\'olica,
     C.P. 38071 BR-22453 Rio de Janeiro, Brazil \\
     \indent~~and Inst. de F\'{\i}sica, Univ. Estadual do Rio de Janeiro,
     rua S\~{a}o Francisco Xavier 524, Rio de Janeiro, Brazil
    \label{BRASIL}}
\titlefoot{Coll\`ege de France, Lab. de Physique Corpusculaire, IN2P3-CNRS,
     FR-75231 Paris Cedex 05, France
    \label{CDF}}
\titlefoot{CERN, CH-1211 Geneva 23, Switzerland
    \label{CERN}}
\titlefoot{Institut de Recherches Subatomiques, IN2P3 - CNRS/ULP - BP20,
     FR-67037 Strasbourg Cedex, France
    \label{CRN}}
\titlefoot{Now at DESY-Zeuthen, Platanenallee 6, D-15735 Zeuthen, Germany
    \label{DESY}}
\titlefoot{Institute of Nuclear Physics, N.C.S.R. Demokritos,
     P.O. Box 60228, GR-15310 Athens, Greece
    \label{DEMOKRITOS}}
\titlefoot{Dipartimento di Fisica, Universit\`a di Genova and INFN,
     Via Dodecaneso 33, IT-16146 Genova, Italy
    \label{GENOVA}}
\titlefoot{Institut des Sciences Nucl\'eaires, IN2P3-CNRS, Universit\'e
     de Grenoble 1, FR-38026 Grenoble Cedex, France
    \label{GRENOBLE}}
\titlefoot{Helsinki Institute of Physics, HIP,
     P.O. Box 9, FI-00014 Helsinki, Finland
    \label{HELSINKI}}
\titlefoot{Joint Institute for Nuclear Research, Dubna, Head Post
     Office, P.O. Box 79, RU-101 000 Moscow, Russian Federation
    \label{JINR}}
\titlefoot{Institut f\"ur Experimentelle Kernphysik,
     Universit\"at Karlsruhe, Postfach 6980, DE-76128 Karlsruhe,
     Germany
    \label{KARLSRUHE}}
\titlefoot{Institute of Nuclear Physics,Ul. Kawiory 26a,
     PL-30055 Krakow, Poland
    \label{KRAKOW1}}
\titlefoot{Faculty of Physics and Nuclear Techniques, University of Mining
     and Metallurgy, PL-30055 Krakow, Poland
    \label{KRAKOW2}}
\titlefoot{Universit\'e de Paris-Sud, Lab. de l'Acc\'el\'erateur
     Lin\'eaire, IN2P3-CNRS, B\^{a}t. 200, FR-91405 Orsay Cedex, France
    \label{LAL}}
\titlefoot{School of Physics and Chemistry, University of Landcaster,
     Lancaster LA1 4YB, UK
    \label{LANCASTER}}
\titlefoot{LIP, IST, FCUL - Av. Elias Garcia, 14-$1^{o}$,
     PT-1000 Lisboa Codex, Portugal
    \label{LIP}}
\titlefoot{Department of Physics, University of Liverpool, P.O.
     Box 147, Liverpool L69 3BX, UK
    \label{LIVERPOOL}}
\titlefoot{LPNHE, IN2P3-CNRS, Univ.~Paris VI et VII, Tour 33 (RdC),
     4 place Jussieu, FR-75252 Paris Cedex 05, France
    \label{LPNHE}}
\titlefoot{Department of Physics, University of Lund,
     S\"olvegatan 14, SE-223 63 Lund, Sweden
    \label{LUND}}
\titlefoot{Universit\'e Claude Bernard de Lyon, IPNL, IN2P3-CNRS,
     FR-69622 Villeurbanne Cedex, France
    \label{LYON}}
\titlefoot{Dipartimento di Fisica, Universit\`a di Milano and INFN-MILANO,
     Via Celoria 16, IT-20133 Milan, Italy
    \label{MILANO}}
\titlefoot{Dipartimento di Fisica, Univ. di Milano-Bicocca and
     INFN-MILANO, Piazza della Scienza 2, IT-20126 Milan, Italy
    \label{MILANO2}}
\titlefoot{IPNP of MFF, Charles Univ., Areal MFF,
     V Holesovickach 2, CZ-180 00, Praha 8, Czech Republic
    \label{NC}}
\titlefoot{NIKHEF, Postbus 41882, NL-1009 DB
     Amsterdam, The Netherlands
    \label{NIKHEF}}
\titlefoot{National Technical University, Physics Department,
     Zografou Campus, GR-15773 Athens, Greece
    \label{NTU-ATHENS}}
\titlefoot{Physics Department, University of Oslo, Blindern,
     NO-0316 Oslo, Norway
    \label{OSLO}}
\titlefoot{Dpto. Fisica, Univ. Oviedo, Avda. Calvo Sotelo
     s/n, ES-33007 Oviedo, Spain
    \label{OVIEDO}}
\titlefoot{Department of Physics, University of Oxford,
     Keble Road, Oxford OX1 3RH, UK
    \label{OXFORD}}
\titlefoot{Dipartimento di Fisica, Universit\`a di Padova and
     INFN, Via Marzolo 8, IT-35131 Padua, Italy
    \label{PADOVA}}
\titlefoot{Rutherford Appleton Laboratory, Chilton, Didcot
     OX11 OQX, UK
    \label{RAL}}
\titlefoot{Dipartimento di Fisica, Universit\`a di Roma II and
     INFN, Tor Vergata, IT-00173 Rome, Italy
    \label{ROMA2}}
\titlefoot{Dipartimento di Fisica, Universit\`a di Roma III and
     INFN, Via della Vasca Navale 84, IT-00146 Rome, Italy
    \label{ROMA3}}
\titlefoot{DAPNIA/Service de Physique des Particules,
     CEA-Saclay, FR-91191 Gif-sur-Yvette Cedex, France
    \label{SACLAY}}
\titlefoot{Instituto de Fisica de Cantabria (CSIC-UC), Avda.
     los Castros s/n, ES-39006 Santander, Spain
    \label{SANTANDER}}
\titlefoot{Inst. for High Energy Physics, Serpukov
     P.O. Box 35, Protvino, (Moscow Region), Russian Federation
    \label{SERPUKHOV}}
\titlefoot{J. Stefan Institute, Jamova 39, SI-1000 Ljubljana, Slovenia
     and Laboratory for Astroparticle Physics,\\
     \indent~~Nova Gorica Polytechnic, Kostanjeviska 16a, SI-5000 Nova Gorica, Slovenia, \\
     \indent~~and Department of Physics, University of Ljubljana,
     SI-1000 Ljubljana, Slovenia
    \label{SLOVENIJA}}
\titlefoot{Fysikum, Stockholm University,
     Box 6730, SE-113 85 Stockholm, Sweden
    \label{STOCKHOLM}}
\titlefoot{Dipartimento di Fisica Sperimentale, Universit\`a di
     Torino and INFN, Via P. Giuria 1, IT-10125 Turin, Italy
    \label{TORINO}}
\titlefoot{Dipartimento di Fisica, Universit\`a di Trieste and
     INFN, Via A. Valerio 2, IT-34127 Trieste, Italy \\
     \indent~~and Istituto di Fisica, Universit\`a di Udine,
     IT-33100 Udine, Italy
    \label{TU}}
\titlefoot{Univ. Federal do Rio de Janeiro, C.P. 68528
     Cidade Univ., Ilha do Fund\~ao
     BR-21945-970 Rio de Janeiro, Brazil
    \label{UFRJ}}
\titlefoot{Department of Radiation Sciences, University of
     Uppsala, P.O. Box 535, SE-751 21 Uppsala, Sweden
    \label{UPPSALA}}
\titlefoot{IFIC, Valencia-CSIC, and D.F.A.M.N., U. de Valencia,
     Avda. Dr. Moliner 50, ES-46100 Burjassot (Valencia), Spain
    \label{VALENCIA}}
\titlefoot{Institut f\"ur Hochenergiephysik, \"Osterr. Akad.
     d. Wissensch., Nikolsdorfergasse 18, AT-1050 Vienna, Austria
    \label{VIENNA}}
\titlefoot{Inst. Nuclear Studies and University of Warsaw, Ul.
     Hoza 69, PL-00681 Warsaw, Poland
    \label{WARSZAWA}}
\titlefoot{Fachbereich Physik, University of Wuppertal, Postfach
     100 127, DE-42097 Wuppertal, Germany
    \label{WUPPERTAL}}
\addtolength{\textheight}{-10mm}
\addtolength{\footskip}{5mm}
\clearpage
\headsep 30.0pt
\end{titlepage}
%
\pagenumbering{arabic} 
\setcounter{footnote}{0} %
\large
\section{Introduction}
An important component of the $e^+e^-$ collisions at LEP II energy
is the two-photon fusion process.
It has been pointed out that two-photon production of inclusive $\jpsi$ mesons:
\bln{jpse}
  e^++e^-\to e^++e^-+\gamma_{_1}+\gamma_{_2},
\eln
\bln{jps0}
  \gamma_{_1}+\gamma_{_2}\to\jpsi+X.
\eln
is a sensitive channel for investigating
the gluon distribution in the photon~\cite{LEP2}.

\begin{figure}[ht]
\hspace{6cm} a) \hfill b) \\
\mbox{\epsfig{file=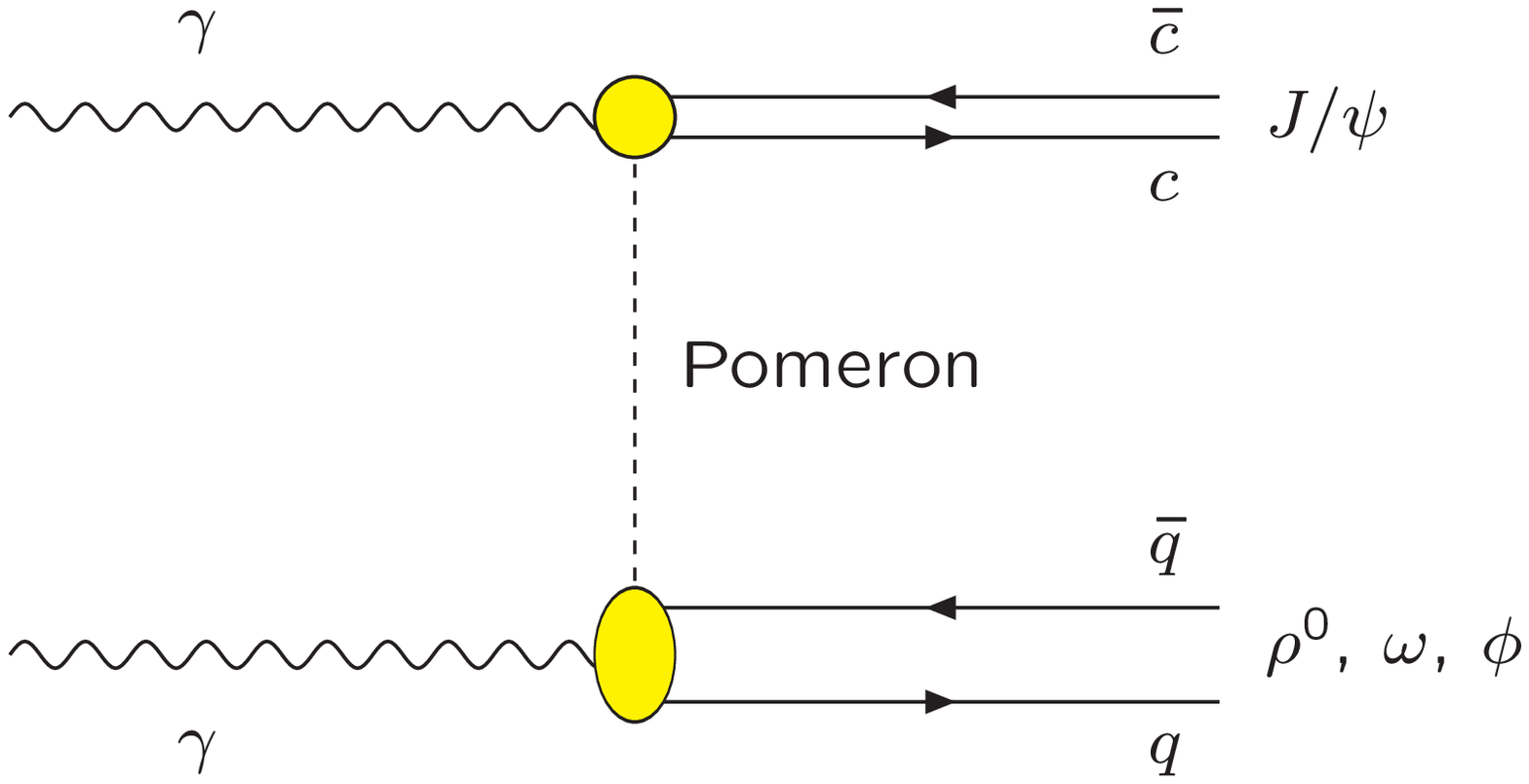,width=7cm}}
\hfill
\mbox{\epsfig{file=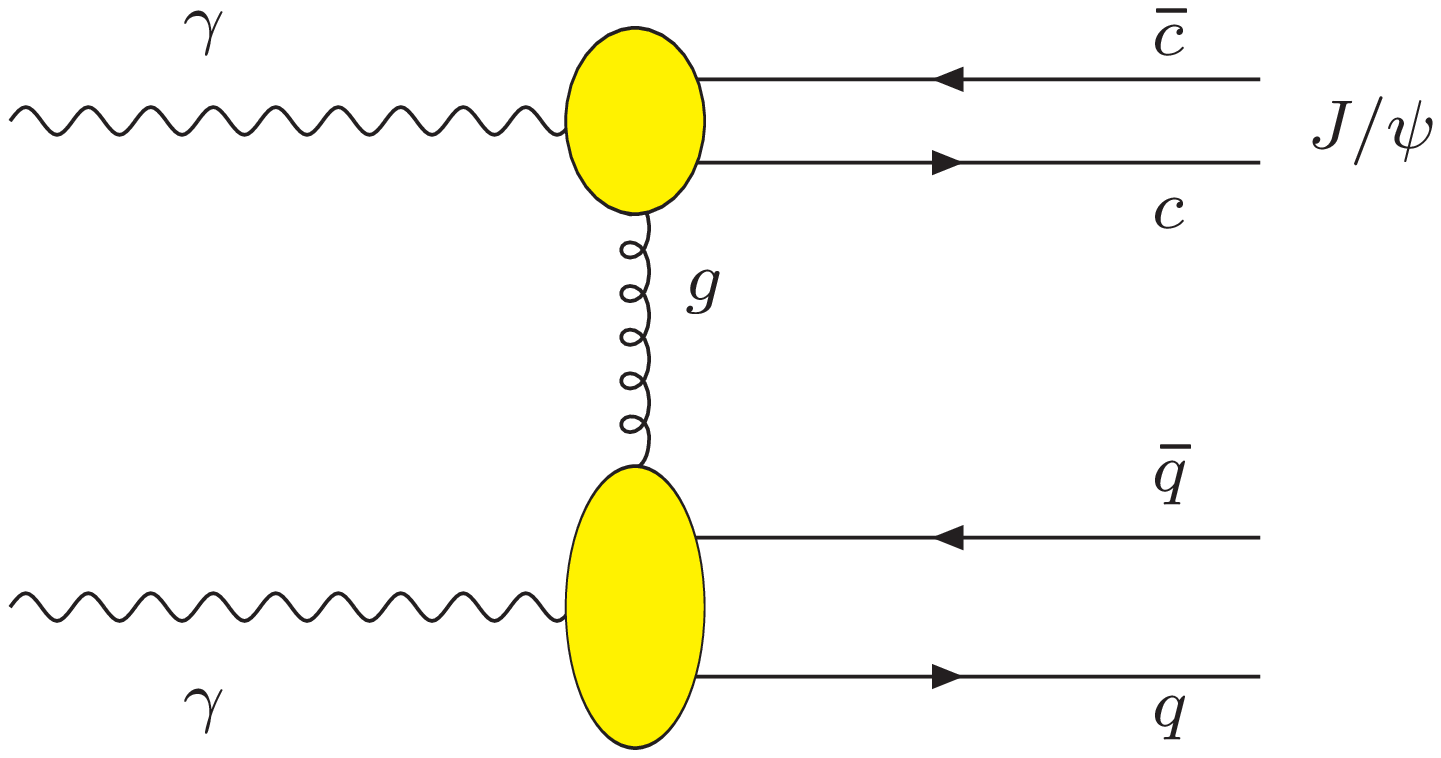,width=7cm}}
\caption{Inclusive $J/\psi$ production in $\gamma\gamma$ processes:
a) through vector-meson dominance, b) via gluon content of the photon,
i.e. `resolved' contributions.}
\label{fg01}
\end{figure}

   There are two important processes leading to inclusive \jpsi\ production.
The corresponding typical diagrams are given in Fig.~\ref{fg01}a--b.
Less important diagrams are not considered here.
The first process is described by
the vector-meson dominance (VDM) model~\cite{VDM}:
\bln{rc1}
  \gamma_{_1}\to c+\bar c,
\qquad\gamma_{_2}\to q+\bar q
\qquad{\rm and}\qquad\gamma_{_1}+\gamma_{_2}\to\jpsi+q+\bar q,
\eln
as shown in Fig.~\ref{fg01}a.  The vertices for $\gamma_{_1}$ and
$\gamma_{_2}$ are connected by Pomeron exchange or diffractive
dissociation of photons.  The final-state parton pairs $c+\bar c$ and
$q+\bar q$ are both in the state of $J^{PC}=1^{--}$, which means
that the latter is dominated by the low-mass vector mesons $\rho^0$,
$\omega$ and $\phi$ but a more general inclusive hadronization of the
partons may also be important.
\noindent

   The second process is described, for example, by the colour-octet
model~\cite{CSM}.
It proceeds through the so-called `resolved' contribution of the
photons, in which the intermediate photons are `resolved' into
their constituent partons:
\bln{rc2}
  \gamma_{_1}+g_{_\gamma}\to c+\bar c,\qquad
 \gamma_{_2}+g_{_\gamma}\to q+\bar q,\qquad{\rm and}\qquad
     \gamma_{_1}+\gamma_{_2}\to\jpsi+q+\bar q,
\eln
as shown in Fig.~\ref{fg01}b.

\noindent
It is seen that this process requires
production of a `resolved' gluon ($g_{_\gamma}$) from both photons.  
Thus, this production mechanism provides a sensitive probe of
the gluon content of the photon.

   The purpose of this letter is to report the observation of inclusive
\jpsi\ production from the two-photon fusion process, to give
its production characteristics along with the cross-section and finally
to assess the relative importance of the 
production processes discussed above.
Section 2 describes the selection criteria for the event sample collected
for this study. The measurement of inclusive \jpsi\ production in the
$\mu^+\mu^-$ channel and its interpretation in terms of diffractive and
resolved  
processes is presented in section 3 followed by a summary and conclusions.
\section{Experimental Procedure}
The analysis presented here is based on the data taken with the DELPHI
detector~\cite{DELPHI1,DELPHI2} during the years 1996--2000, 
excluding the part of the data collected in the last period of 2000, 
when one of the Time Projection Chamber (TPC) sectors was not in operation.
The centre-of-mass energies
$\sqrt{s}$ for LEP ranged from 161 to 207 GeV. 
The total integrated luminosity
used in the analysis is 617 pb$^{-1}$.
 

  The charged particle tracks were measured in the 1.2 T magnetic field by a
set of tracking detectors including 
the microVertex Detector (VD), the Inner Detector (ID),
the TPC, the Outer Detector (OD) 
and the Forward/Backward Chambers
FCA and FCB. The following selection criteria were applied:
\begin{list}{}{\setlength{\leftmargin}{42pt}
        \setlength{\topsep}{0pt}
                    \setlength{\itemsep}{0pt}
                         \setlength{\parsep}{0pt}}
\item[($a$)] particle momentum $p \ >$ 200 \mevc;
\item[($b$)] relative momentum error of a track $\Delta p/p \ <$ 100\%;
\item[($c$)] impact parameter of a track, transverse to the beam axis 
$\ <$ 3 cm;
\item[($d$)] impact parameter of a track, along the beam axis $\ <$ 7 cm;
\item[($e$)] polar angle of a track, with respect to the beam axis
              $10^{\circ}~<~\theta~<~170^{\circ}$;
\item[($f$)] track length $\ >$ 30 cm.
\end{list}
The neutral particles ($\gamma$, $\pi^0$, ${K}^0_{_L}$, $n$) 
were selected by demanding that the calorimetric
information, not associated with charged particle tracks, 
satisfies the following cuts:
\begin{list}{}{\setlength{\leftmargin}{42pt}
        \setlength{\topsep}{0pt}
                    \setlength{\itemsep}{0pt}
                         \setlength{\parsep}{0pt}}
\item[($g$)] $E(\rm neutral)>0.2$ GeV for the electromagnetic showers, 
          unambiguously identified as photons;
\item[($h$)] $E(\rm neutral)>0.5$ GeV for all the other showers;
\item[($i$)] polar angle of neutral particle tracks, with respect 
to the beam axis
              $10^{\circ}~<~\theta~<~170^{\circ}$.
\end{list}

   In order to ensure a very high trigger efficiency,
the selected events were required to satisfy at least
{\em one} of the following sets of criteria:
\begin{list}{}{\setlength{\leftmargin}{42pt}
        \setlength{\topsep}{0pt}
                    \setlength{\itemsep}{0pt}
                         \setlength{\parsep}{0pt}}
\item[($j_1$)] one or more charged particle tracks in the barrel region 
($40^{\circ} < \theta < 140^{\circ}$) with $p_{t}>1.2\,\gevc$, is found;
\item[($j_2$)] one or more neutral particle tracks in the Forward 
ElectroMagnetic
 Calorimeter (FEMC) ( $10^{\circ} < \theta < 36^{\circ}$ and 
$144^{\circ} < \theta < 170^{\circ}$)
with energy greater than 10~GeV, is found;
\item[($j_3$)] the total sum of charged particle tracks in the barrel 
with $p_{t} > 1\,\gevc$, of charged particle tracks in the forward region 
($10^{\circ} < \theta < 40^{\circ}$ or
$140^{\circ} < \theta < 170^{\circ}$)
with $p_{t} > 2\,\gevc$ and of neutral particle tracks in the 
FEMC with $E > 7$ GeV, is greater than one;
\item[($j_4$)] the total sum of charged particle tracks 
in the barrel with $p_{t}>0.5 \,\gevc$,
of charged particle tracks in the forward region with $p_{t} > 1\,\gevc$ and
of neutral particle tracks in the FEMC with $E > 5$ GeV, 
is greater than four. 
\end{list}
The trigger efficiency for the events which passed 
the above requirements is bigger than 98\%.   

\begin{figure}[ht]
\begin{center}\mbox{
\epsfig{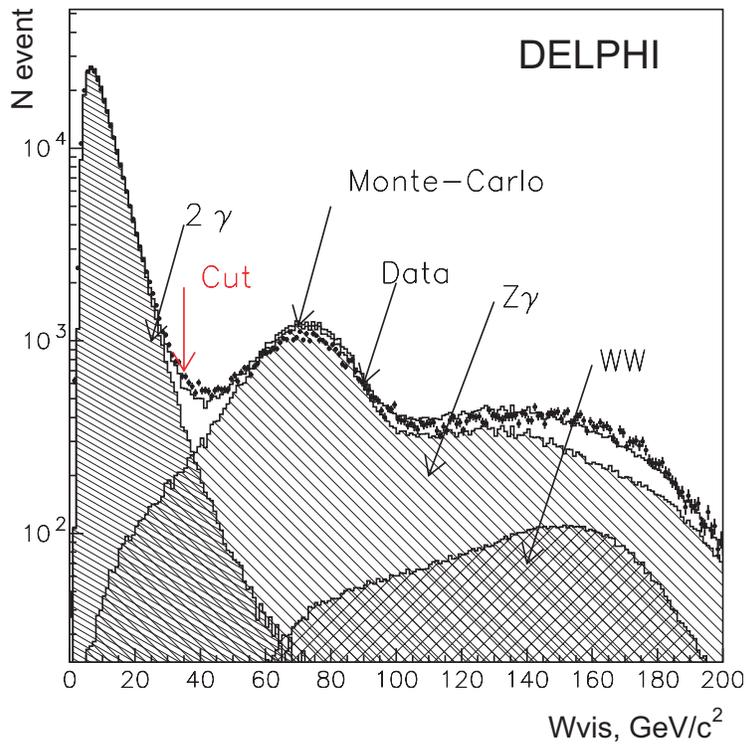}
}\end{center}
\vskip-4mm
\caption{$W_{\rm vis}$ distributions  for the LEP~II DELPHI data,
for the simulated $\gamma\gamma\rightarrow \ {\rm hadrons}$,
${e}^+{e}^-\rightarrow {Z}^0 \gamma$,
${e}^+{e}^-\rightarrow {W}^+ {W}^-$ and
the sum of all above Monte-Carlo contributions}
\label{fg02}
\end{figure}

    The hadronic two-photon events are characterized by a
low visible invariant mass. 
Consequently, the following additional cuts were applied:
\begin{list}{}{\setlength{\leftmargin}{42pt}
        \setlength{\topsep}{0pt}
                    \setlength{\itemsep}{0pt}
                         \setlength{\parsep}{0pt}}
\item[($k$)] the visible invariant mass, $W_{\rm vis}$,
calculated from the four-momentum vectors of the
measured charged and neutral particle tracks, 
is less than  35 \gevcc;
\item[($l$)] the number of charged particle tracks $N_{\rm ch}$ 
satisfies $4 \ \leq \ N_{\rm ch} \ \leq \ 30$;
\item[($m$)] the sum of the transverse energy components with respect
to the beam direction for all charged
particle tracks ($\sum{\sqrt{p_t^2+m^2_\pi}}$) is greater than 3 GeV.
\end{list}
The comparison  of the $W_{\rm vis}$ distributions, after
the cuts on $N_{\rm ch}$ and 
$\sum{E^{\rm vis}_{_T}}$, both for the data and
the events simulated by PYTHIA,
shows (Fig.~\ref{fg02}) that the cut
$W_{\rm vis} \ \leq$ 35 \gevcc\
rejects the major part of the non-two-photon events.

 A total of $N_t=274\,510$ events
remain in the data sample after applying all these cuts.
The main background comes from the process
${e}^+{e}^-\rightarrow Z \gamma$
and amounts to $\sim$1.2\% of the selected $\gamma\gamma$ events.
The background from the ${e}^+{e}^-\rightarrow {W}^+{W}^-$ is negligible, 
as seen in Fig.~\ref{fg02}.

  $\jpsi$ candidates have been selected
using the $\mu^+\mu^-$ decay channel.  
For the muon pair selection, the following criteria were imposed:
\begin{list}{}{\setlength{\leftmargin}{42pt}
        \setlength{\topsep}{0pt}
                    \setlength{\itemsep}{0pt}
                         \setlength{\parsep}{0pt}}
\item[($n$)]  at least two charged particle tracks, with zero net charge, 
should be accepted by the standard DELPHI 
muon-tagging algorithm~\cite{DELPHI2},
or be identified as muons by the hadronic calorimeter;
\item[($o$)] the tracks should not come from any reconstructed secondary vertex
or be \mbox{identified} as a kaon, proton or electron by the standard DELPHI
identification packages. 
\end{list}

\section{Inclusive \jpsi\ Production}

In this section, we first determine the inclusive $\jpsi$ production in the
$\mu^+\mu^-$ channel. Then we interpret it in terms of diffractive and resolved
processes by fitting the experimental $p_{_T}^2(\jpsi)$
 distribution to the PYTHIA
predictions for both processes. This allows to deduce the cross-section for
inclusive $\jpsi$ production, taking into account the 
$\gamma\,\gamma\to\jpsi+X$
and the $\jpsi\to\mu^+\mu^-$ efficiencies.
 As the first set of efficiencies is
model-dependent, we also give the `visible' cross-section in which only
the detector efficiency for $\jpsi\to\mu^+\mu^-$ decay is considered. 
We finally present the $\jpsi$ production characteristics together with 
the PYTHIA predictions.

\begin{figure}[ht]\vskip2mm
\begin{center}\mbox{
\epsfig{file=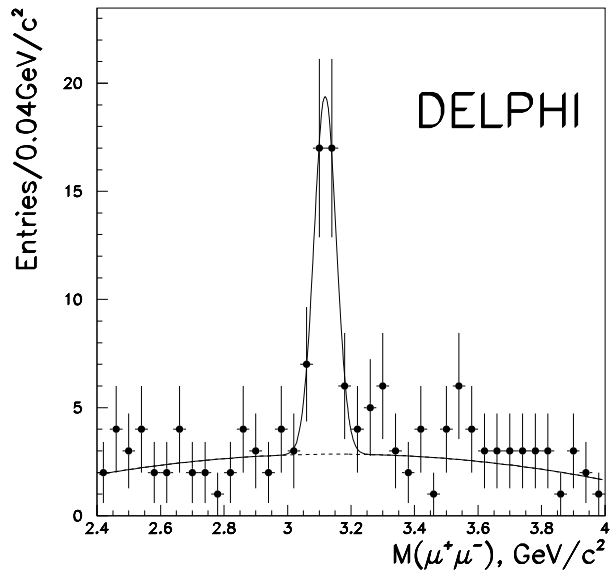,width=8.0cm}
}\end{center}
\vskip 3mm
\caption{$M(\mu^+\mu^-)$ distribution from the LEP~II DELPHI data.
The solid curve corresponds to a Gaussian fit over a second order polynomial
background.}
\label{fg03}
\vskip2mm
\end{figure}

In Fig.~\ref{fg03} we give the invariant mass distribution for 
identified $\mu^+\mu^-$ pairs, selected as outlined in the previous section.
The \jpsi\ signal shows up over little background.
A least squares fit to the $M(\mu^+\mu^-)$ distribution with a Gaussian
for the signal and a second order polynomial for the background
gives the following results:
\vskip8pt\begin{center}\begin{tabular}{lc@{$\ =\ $}r@{$\pm$}ll}
\jpsi\ mass:&$M$&3119&8&\mevcc,\\
\jpsi\ width:&$\sigma({\rm obs})$&35&7&\mevcc.
\end{tabular}\end{center}\vskip8pt

\vskip7mm
The observed width of the peak is consistent within errors with
the invariant mass resolution of a pair of charged particle tracks in the
mass region around 3\,\gevcc. The number of observed events from the fit is:
\begin{center}\begin{tabular}{lc@{$\ =\ $}r@{$\pm$}ll}
 &$N$(\jpsi)&36&7&events,
\end{tabular}\end{center}
over a background of about 11 events. 

If we take the L3 result~\cite{L3} for the 
beauty cross-section from $\gamma\gamma$ events and the PDG value~\cite{PDG}
for the branching ratio of beauty hadrons to \jpsi, the expected number of
\mbox{$\jpsi\to\mu^+\mu^-$} from beauty hadrons is $2.1\pm0.6$.  The backgrounds
from the processes $e^+ + e^- \to Z+\gamma\to\jpsi+X$ and
$\gamma+\gamma\to \chi_{c2}\to\jpsi+\pi^++\pi^-+\pi^0$ are less than
0.20 and 0.30 event respectively. According to the selection criteria the 
system X
contains at least two charged particle tracks, hence we do not consider
such sources of \jpsi~ production as 
$\gamma+\gamma\to \chi_{c2}\to\jpsi+\gamma$.
We checked that in the four-prong events with $\jpsi\to\mu^+\mu^-$ candidates
there are no photon conversions.

We used the PYTHIA 6.156 generator~\cite{PYTHIA} to estimate the efficiency.
The generated events were passed through the simulation package of
the DELPHI detector~\cite{DELPHI2} and then processed with the same
reconstruction and analysis programs as the real data.
There is a substantial fraction of PYTHIA events where $\jpsi$ mesons are
produced just as a simple fusion of two photons because there is not enough
phase space to produce additional particles.
We do not use such events. 
The process where both photons are VDM photons we
will call `diffractive' and the process without VDM photons we will
call `resolved'.

\begin{figure}[ht]\vskip-6mm
\begin{center}
\mbox{\epsfig{file=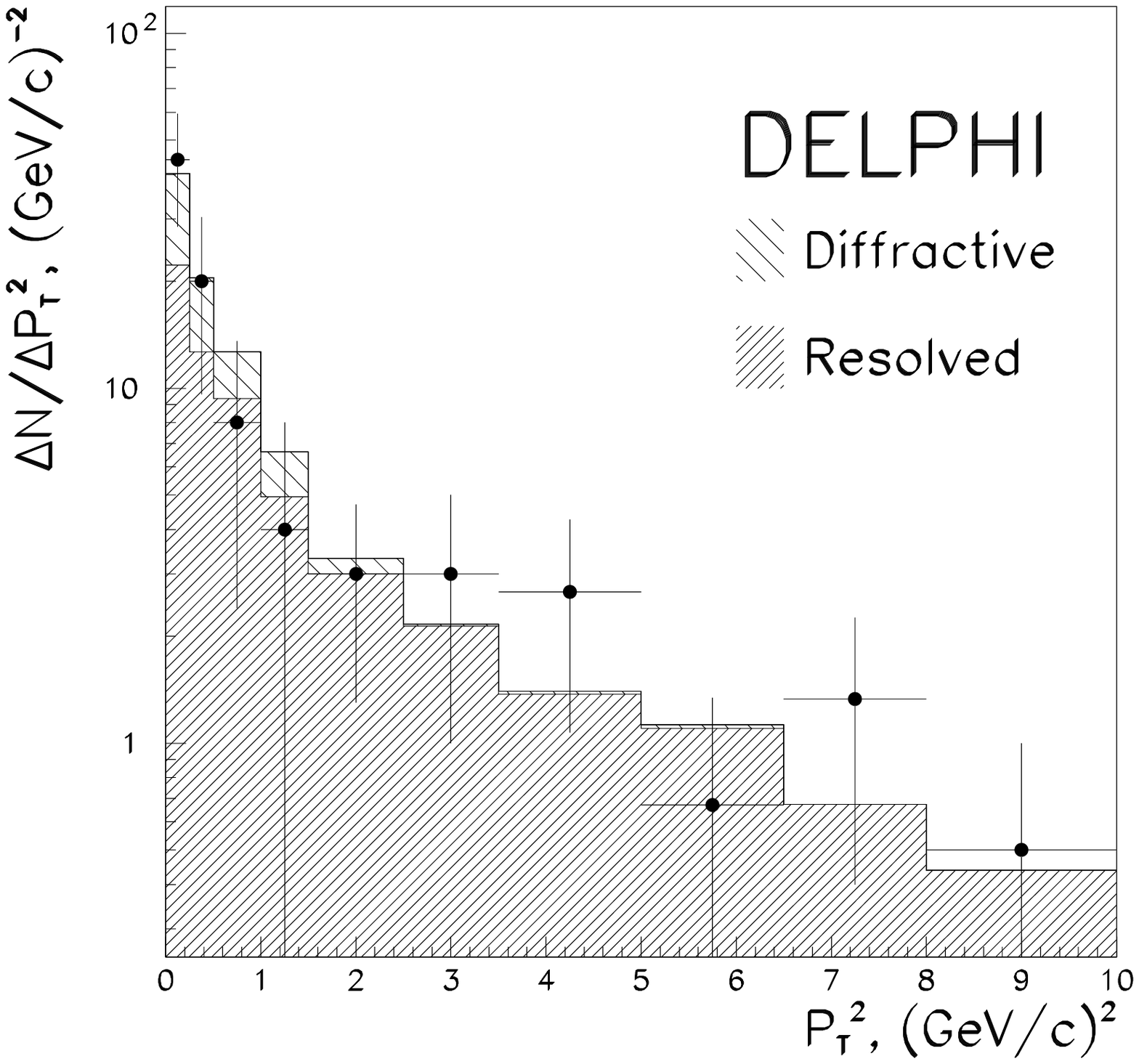,width=8.0cm}}
\end{center}
\vskip-3mm
\caption{
$p_{_T}^2(\jpsi)$ distribution from the LEP~II DELPHI data, shown as
points with error bars.  The histogram is a combination of the normalized 
`resolved' and `diffractive' processes from PYTHIA (see text).}
\label{fg04}
\end{figure}


A set of the \jpsi~ production characteristics is exhibited
in Figs.4 to 8. For each bin of every distribution shown, 
we have examined the $M(\mu^+\mu^-)$ spectrum and then
fitted with a Gaussian and a second order polynomial background, 
to get the number of signal events per bin. This
number is then renormalized for each distribution,
so that the total sum is always equal to 36 events. Hence,
Fig.4 to Fig.8 are background subtracted distributions.

\begin{figure}[ht]\vskip-6mm
\begin{center}
\mbox{\epsfig{file=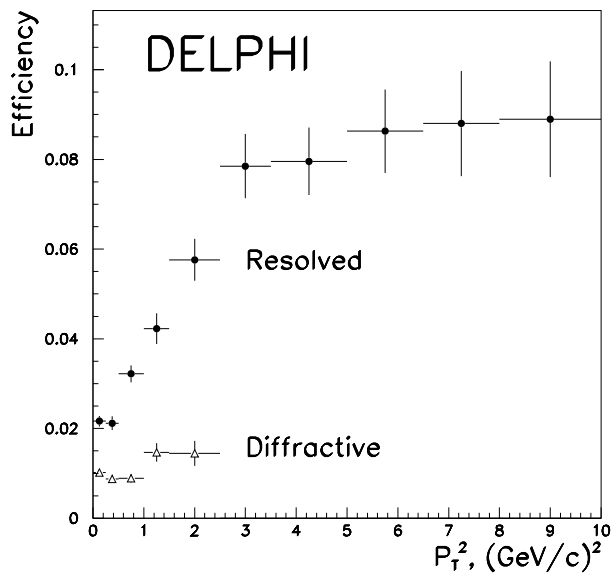,width=8.0cm}}
\end{center}
\vskip-3mm
\caption{
Efficiencies for resolved and diffractive processes 
as functions of $p_{_T}^2$.
}
\label{fg05}
\end{figure}

   Fig.~\ref{fg04} shows the $p_{_T}^2(\jpsi)$ distribution.
As expected, the PYTHIA Monte Carlo prediction for
the $p_{_T}^2(\jpsi)$ distribution is more sharply peaked
near zero for the `diffractive' events (see Fig.~\ref{fg01}a)
than for the `resolved' events (see Fig.~\ref{fg01}b).
We fitted the experimental $p_{_T}^2(\jpsi)$ 
distribution as a function of the two categories of PYTHIA events:
\bln{pta}
   {{\rm d}N\over{\rm d}p_{_T}^2}
 =f\cdot\left.{{\rm d}N\over{\rm d}p_{_T}^2}\right|_{\rm Diffractive}
 +(1-f)\cdot\left.{{\rm d}N\over{\rm d}p_{_T}^2}\right|_{\rm Resolved},
\eln
which gives $f=(26\pm22)\%$ (PYTHIA distributions in Fig.~\ref{fg04} 
are normalized to the data).  
The PYTHIA study tells us that
the experimental efficiencies are very different for the two
categories:
\begin{eqnarray}
 \epsilon({\rm diffractive})&=(0.98\pm0.04)\%,\cr 
 \epsilon({\rm resolved})&=(3.87\pm0.09)\%.
\label{ptb}
\end{eqnarray}
\noindent
According to PYTHIA, about one-half of all the $\gamma\gamma$ events with
$\jpsi\to\mu^+\mu^-$ are produced with the charged particle tracks at polar
angles below 10 degrees, so that they are invisible to the DELPHI detector.
The individual efficiencies as a function of $p_{_T}^2$ are given
in Fig.~\ref{fg05}. Some insight may be gained into these efficiencies
if they are broken down into products of two factors, as follows:
\begin{eqnarray}
 \epsilon({\rm diffractive})
         &=\epsilon_{\gamma\,\gamma}({\rm diffractive})
        \times\epsilon_{\jpsi\to\mu^+\mu^-}({\rm diffractive}),\cr
 \epsilon({\rm resolved})
         &=\epsilon_{\gamma\,\gamma}({\rm resolved})
               \times\epsilon_{\jpsi\to\mu^+\mu^-}({\rm resolved}),
\end{eqnarray}
\noindent
where $\epsilon_{\gamma\,\gamma}$ is the efficiency for
the process $\gamma\,\gamma\to\jpsi+X$ and $\epsilon_{\jpsi\to\mu^+\mu^-}$
is that for $\jpsi\to\mu^+\mu^-$.  As expected, the latter is relatively
process-independent:
\begin{eqnarray}
\epsilon_{\jpsi\to\mu^+\mu^-}({\rm diffractive})
            &=(37.0\pm1.5)\%,\cr
\epsilon_{\jpsi\to\mu^+\mu^-}({\rm resolved})
           &=(32.1\pm0.7)\%.
\end{eqnarray}
\noindent
It is clear, therefore, that the difference in efficiency in (\ref{ptb})
is mostly due to $\epsilon_{\gamma\,\gamma}$. This is highly
process-dependent and hence model-dependent.

   The overall experimental efficiency is:
\bln{ef0}
{1\over\epsilon}={f\over\epsilon({\rm diffractive})}
                     +{1-f\over\epsilon({\rm resolved})},
\eln
which gives $\epsilon=(2.19^{+1.27}_{-0.59})\%$.
Under the assumption that PYTHIA captures the \mbox{kinematical}
features of the resolved and diffractive processes, but not
their absolute cross-sections, the cross-section 
for inclusive $\jpsi$ production is: 
\bln{sg0}
\sigma=N(\jpsi)\cdot(Br\cdot {\cal L}\cdot\epsilon)^{-1} 
               = 45\pm9{\rm (stat)}\pm17{\rm (syst)}\ {\rm pb,}
\eln	       
where $Br=(5.88\pm0.10)\%$ is the branching ratio
for $\jpsi\to\mu^+\mu^-$~\cite{PDG} and
${\cal L}=617\,{\rm pb}^{-1}$ is the total integrated luminosity.
The systematic uncertainties include both the efficiency \eqn{ef0} and the 
branching ratio contributions but not those inherent to the PYTHIA program.

Because of the model-dependent aspect of this analysis
(see, for example, the efficiencies given in (\ref{ptb})),
it is of interest to quote the `visible' cross-section.
Substituting $\epsilon_{\jpsi\to\mu^+\mu^-}({\rm diffractive})$ 
and $\epsilon_{\jpsi\to\mu^+\mu^-}({\rm resolved})$ for 
$\epsilon({\rm diffractive})$ and 
$\epsilon({\rm resolved})$ respectively in \eqn{ef0}, 
the `visible' cross-section can be calculated;
it is: 
\bln{sg1}
\sigma_{\rm vis}=3.0\pm0.6{\rm (stat)}\pm0.1{\rm (syst)}\ {\rm pb.}
\eln
The main source of systematic uncertainty comes from the determination 
of the relative
fractions of resolved and diffractive events which have different
efficiencies (8).
Following the same argument,
we also give the `visible' production rate $\bra n\ket$
for $\jpsi$ production:
\bln{rt0}
  \bra n\ket
=N(\jpsi)\cdot
          \left(N_t\cdot Br\cdot\epsilon_{\jpsi\to\mu^+\mu^-}\right)^{-1} 
               =(6.7\pm1.3{\rm (stat)}\pm 0.3{\rm (syst)})\times10^{-3},
\eln
where $N_t$ is the data sample for the $\gamma\gamma$ selection
as given in the previous section.
 
\begin{figure}[ht]\vskip-6mm
\begin{center}
\mbox{\epsfig{file=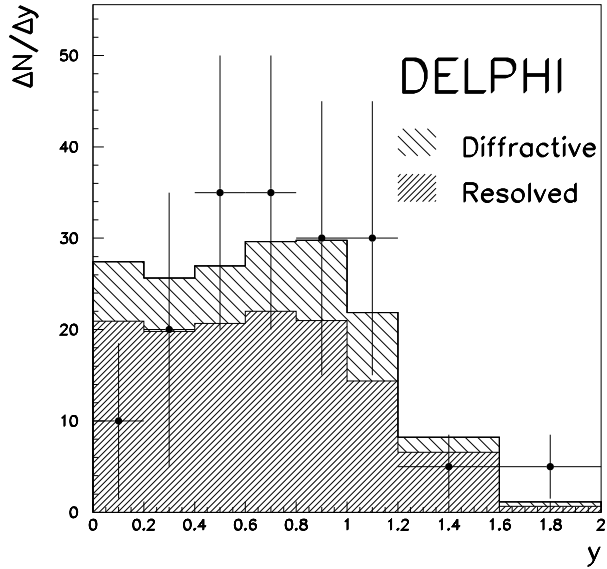,width=8.0cm}}
\end{center}
\vskip-3mm
\caption{
$|y|$ distribution for \jpsi\ mesons from the LEP~II DELPHI data, shown as
points with error bars.  The histogram is a combination of
the normalized `resolved' and `diffractive' processes from PYTHIA (see text).}
\label{fg06}
\end{figure}

\begin{figure}[ht]
\noindent\mbox{\epsfig{file=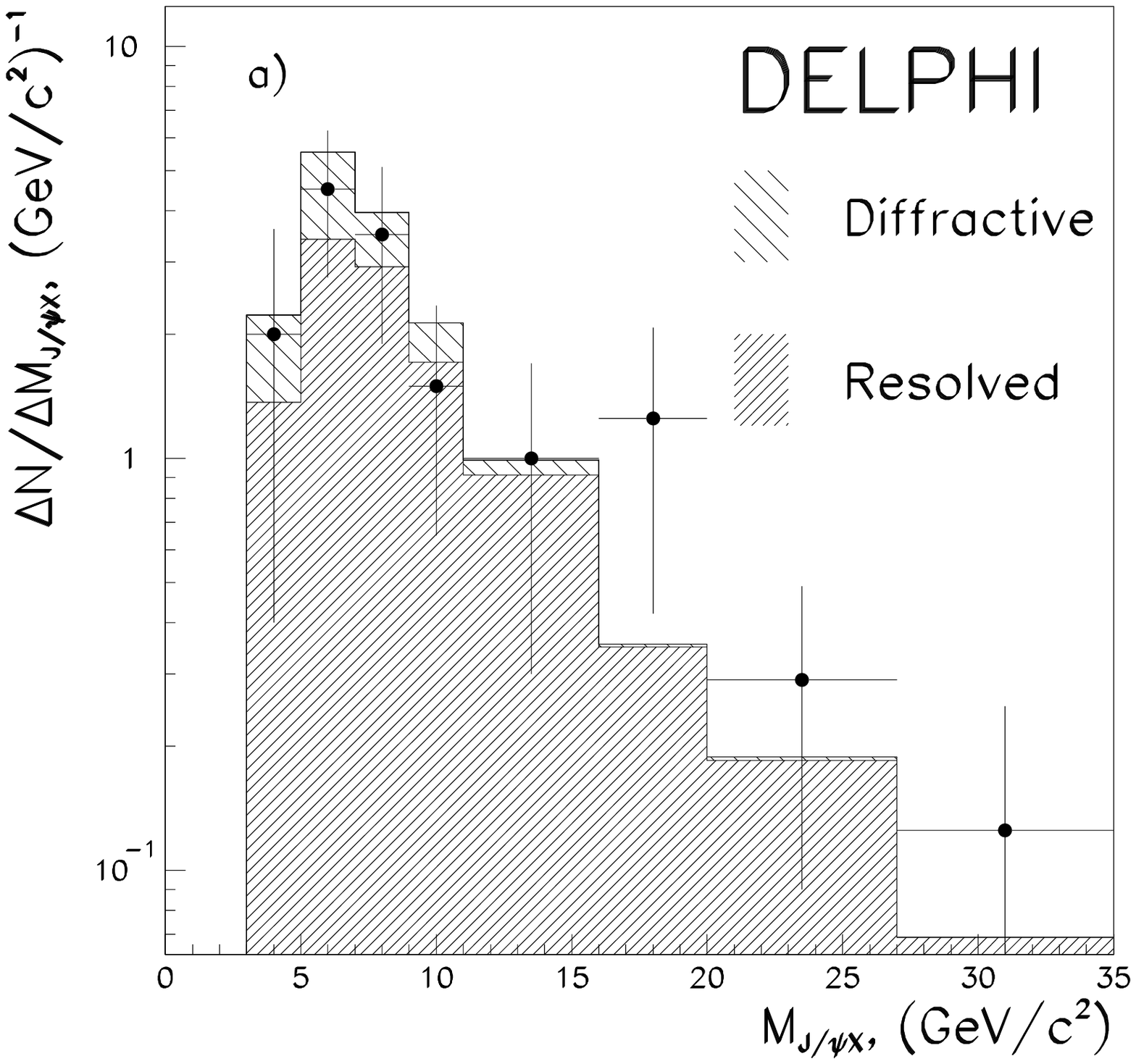,width=7.0cm}}
\hfill
\mbox{\epsfig{file=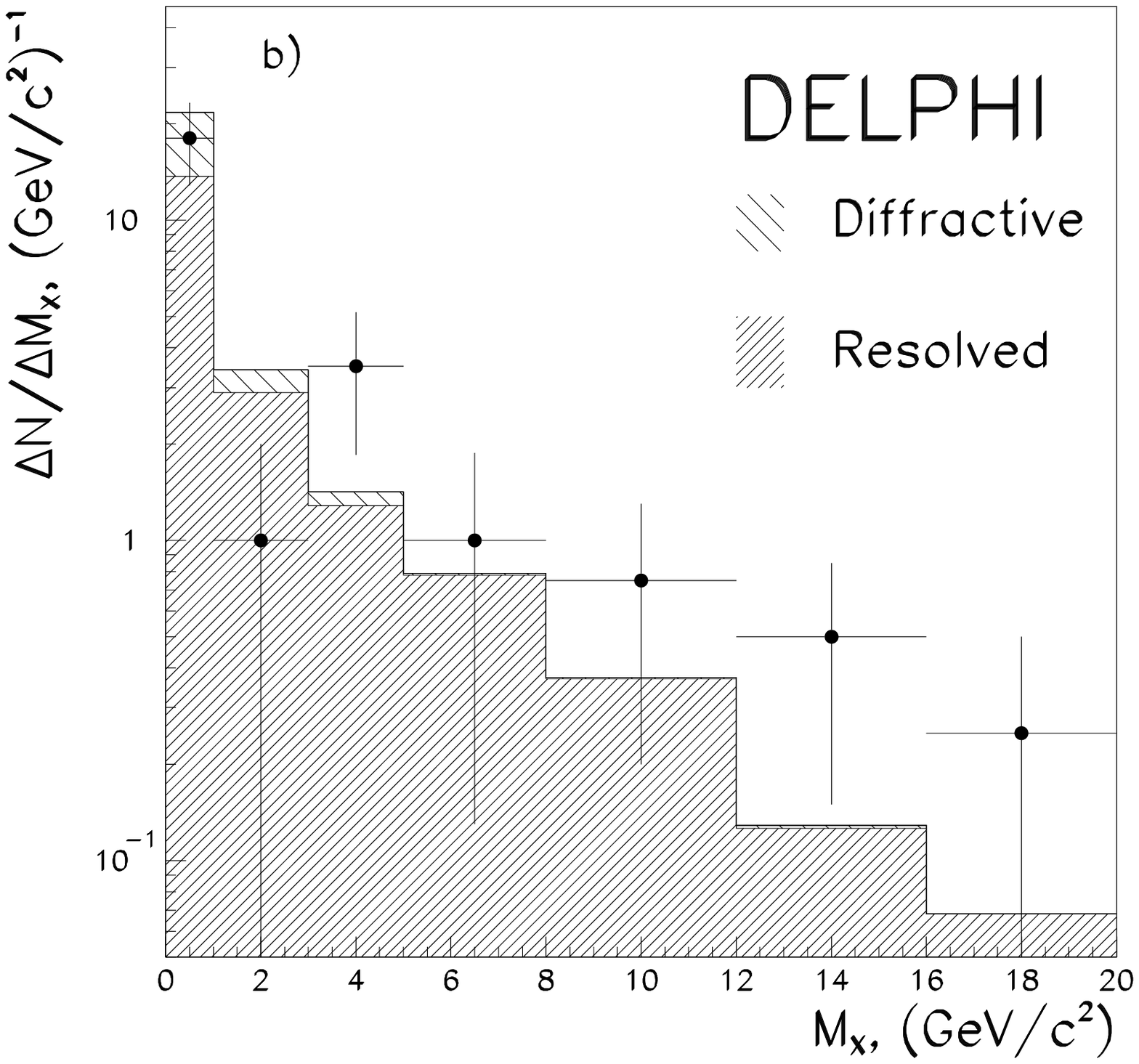,width=7.0cm}} \\
\mbox{\epsfig{file=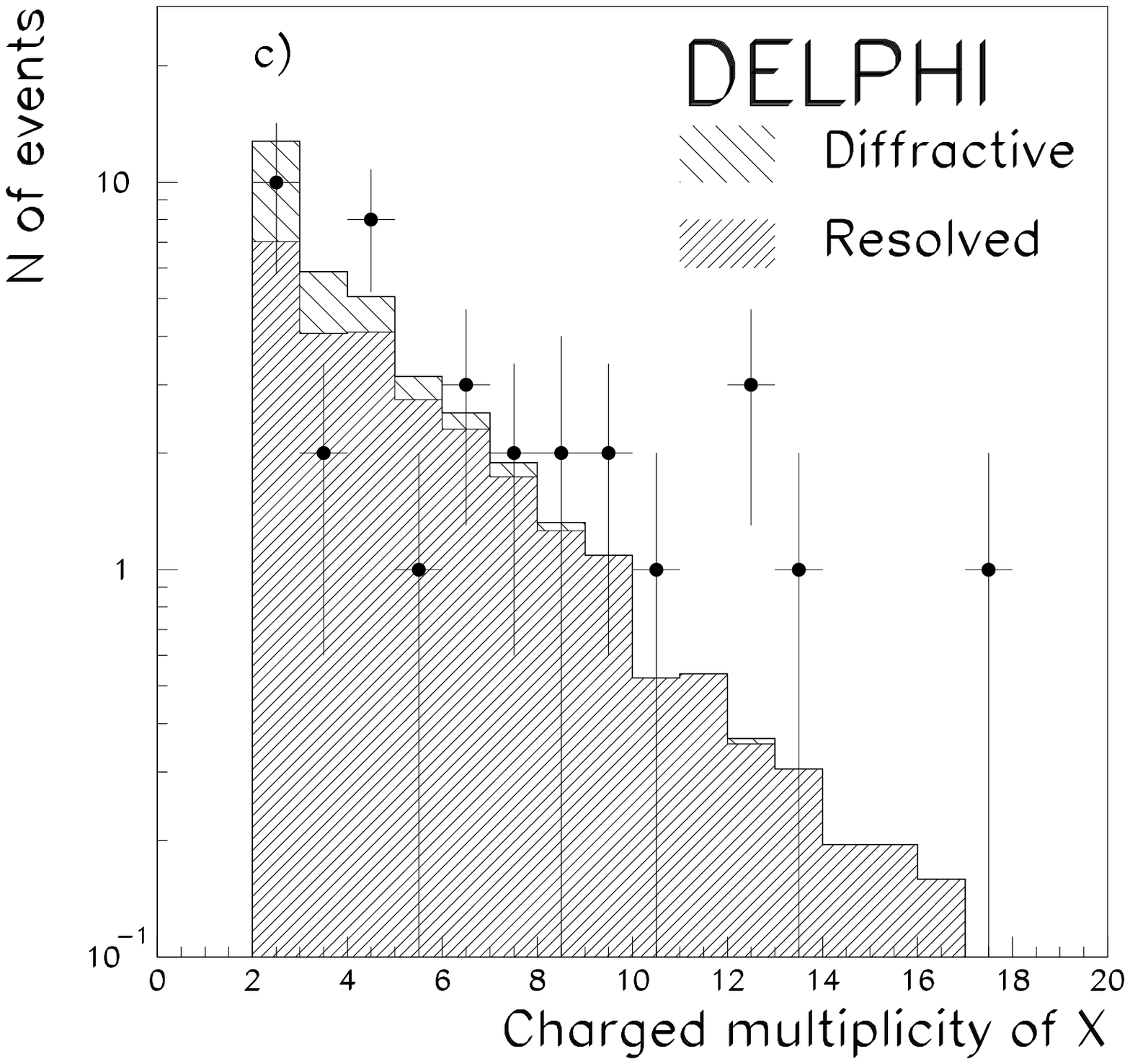,width=7.0cm}}
\hfill
\mbox{\epsfig{file=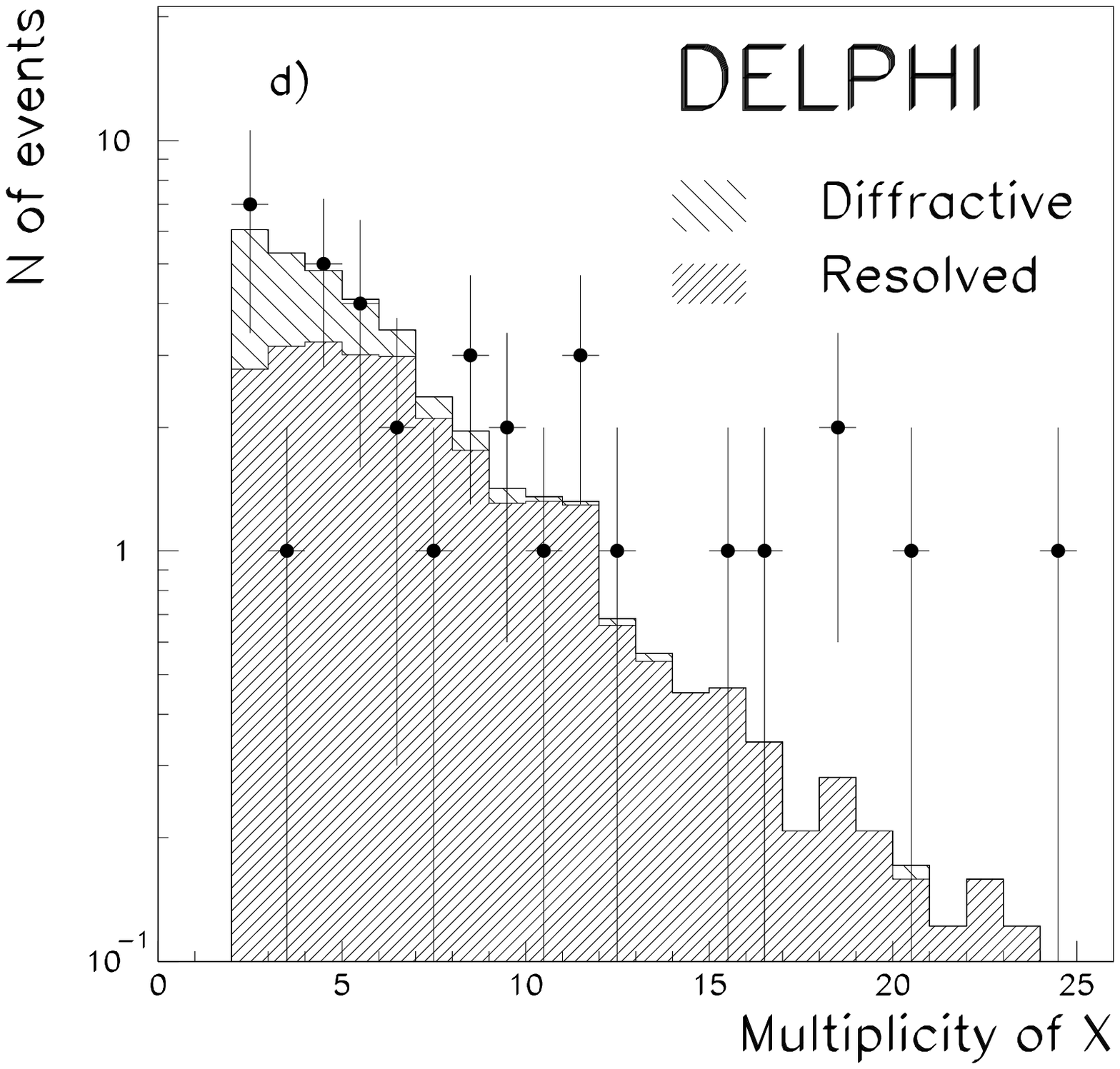,width=7.0cm}}
\caption{Visible distributions of $M(\jpsi + X)$ (a), $M(X)$ 
(b), charged (c) and total (d) multiplicities of the X system. Each
histogram is a combination of the
normalized `resolved' and `diffractive' processes from PYTHIA (see text).}

\label{fg07}
\end{figure}

   The rapidity distribution in the laboratory system
for the \jpsi\ mesons is shown
in Fig.~\ref{fg06}.  The PYTHIA events have been combined
using the same fraction $f$ found in \eqn{pta} and then 
normalized to the observed number of events in $0<|y|<2.0$.
The same techniques have been used to compare the experimental
distributions of $M(\jpsi+X)$, $M(X)$, 
the charged and total multiplicities [$N_{\rm ch}(X)$ and
 $N_{\rm tot}(X)$], in Figs.~\ref{fg07}a--d. 
There is fair agreement within statistics 
between the shapes of our measured distributions
and the PYTHIA predictions (using the best fit as found in \eqn{pta} for the
relative content 
of diffractive and resolved events and renormalizing the PYTHIA
prediction to the number of observed events).

\begin{figure}[ht]\vskip-12mm
\begin{center}
\mbox{\epsfig{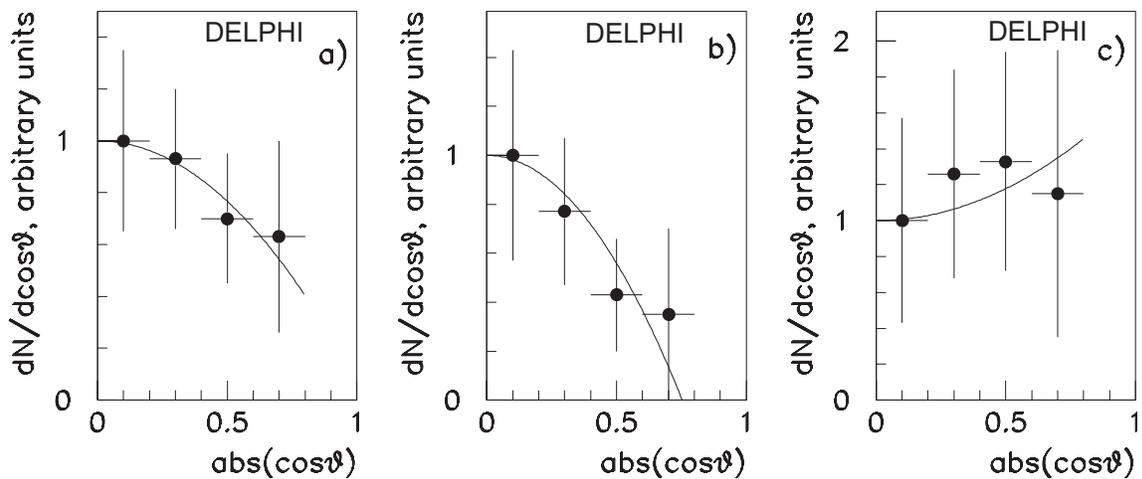}}
\end{center}
\vskip-3mm
\caption{
Acceptance-corrected distributions in $\cos\theta$ where $\theta$
is the helicity angle of $\mu^+$ in the rest frame of $\jpsi\to\mu^+\mu^-$.
The figures (a--c) correspond to the total sample (a), the 
subsamples with $p_{_T}^2(\jpsi)<1.0$ $(\gevc)^2$\ (b) 
and $p_{_T}^2(\jpsi)>1.0$ $(\gevc)^2$\ (c).}
\label{fg08}
\end{figure}

  The acceptance-corrected distributions
in $\cos\theta$, where $\theta$ is the helicity angle of $\mu^+$ 
in the rest frame of $\jpsi\to\mu^+\mu^-$, are shown in Figs.\ref{fg08}a--c,
along with the results of a fit
to the form ($1+a\,\cos^2\theta$).  The fitted parameters are
$a=-0.9\pm0.6$ for the total sample (a), $a=-1.8\pm0.5$ for 
$p_{_T}^2(\jpsi)<1.0$ $(\gevc^2)$ (b) and $a=0.7\pm1.3$ for 
$p_{_T}^2(\jpsi)>1.0$ $(\gevc)^2$~(c).  These results indicate that the \jpsi\
mesons 
are produced with little polarization at high $p_{_T}^2(\jpsi)$, where the
main contribution comes from the resolved processes.
\clearpage
\section{Conclusions}

We have studied the inclusive \jpsi\ production from $\gamma\gamma$
collisions.  The data have been taken by the DELPHI Collaboration
during the LEP II phase, i.e. $\sqrt{s}$ of the LEP machine
ranged from 161 to 207 GeV.  A clear signal
from the reaction $\gamma\gamma\to\jpsi+X$ is seen.

The inclusive cross-section is estimated to be
$\sigma(\jpsi+X)=45\pm9{\rm (stat)}\pm17{\rm (syst)}\ {\rm pb.}$
Based on a study of the $p_T^2$ distribution of different types 
of $\gamma\gamma$
processes in the PYTHIA program, we conclude that
some $(74\pm22\,)$\% of the observed \jpsi\ events are due to the
`resolved' photons, the dominant contribution of which should correspond to 
the gluon content of the photon~\cite{CSM}.

The distributions in $p_{_T}^2(\jpsi)$, $y$ and $\cos\theta$ 
(for $\mu^+$ in the rest frame of $\jpsi\to\mu^+\mu^-$) are presented.
In addition, a study is given of the characteristics of the
system $X$.
All \mbox{distributions} appear to be well reproduced within statistics by the
normalized \mbox{combination} of the fitted `resolved' and `diffractive'
contributions.


\subsection*{Acknowledgements}
\vskip 3 mm
 We are greatly indebted to our technical 
collaborators, to the members of the CERN-SL Division for the excellent 
performance of the LEP collider, and to the funding agencies for their
support in building and operating the DELPHI detector.\\
We acknowledge in particular the support of \\
Austrian Federal Ministry of Education, Science and Culture,
GZ 616.364/2-III/2a/98, \\
FNRS--FWO, Flanders Institute to encourage scientific and technological 
research in the industry (IWT), Federal Office for Scientific, Technical
and Cultural affairs (OSTC), Belgium,  \\
FINEP, CNPq, CAPES, FUJB and FAPERJ, Brazil, \\
Czech Ministry of Industry and Trade, GA CR 202/99/1362,\\
Commission of the European Communities (DG XII), \\
Direction des Sciences de la Mati$\grave{\mbox{\rm e}}$re, CEA, France, \\
Bundesministerium f$\ddot{\mbox{\rm u}}$r Bildung, Wissenschaft, Forschung 
und Technologie, Germany,\\
General Secretariat for Research and Technology, Greece, \\
National Science Foundation (NWO) and Foundation for Research on Matter (FOM),
The Netherlands, \\
Norwegian Research Council,  \\
State Committee for Scientific Research, Poland, SPUB-M/CERN/PO3/DZ296/2000,
SPUB-M/CERN/PO3/DZ297/2000 and 2P03B 104 19 and 2P03B 69 23(2002-2004)\\
JNICT--Junta Nacional de Investiga\c{c}\~{a}o Cient\'{\i}fica 
e Tecnol$\acute{\mbox{\rm o}}$gica, Portugal, \\
Vedecka grantova agentura MS SR, Slovakia, Nr. 95/5195/134, \\
Ministry of Science and Technology of the Republic of Slovenia, \\
CICYT, Spain, AEN99-0950 and AEN99-0761,  \\
The Swedish Natural Science Research Council,      \\
Particle Physics and Astronomy Research Council, UK, \\
Department of Energy, USA, DE-FG02-01ER41155, \\
EEC RTN contract HPRN-CT-00292-2002. \\
We are indebted to Dr. T. Sj\"ostrand for his help with
the diagrams included in this paper and for his comments on PYTHIA.
We thank M. Klasen for his help with the diagrams important
for inclusive \jpsi\ production.
S.-U. Chung is grateful for the warm hospitality extended to him
by the CERN staff during his sabbatical year in the EP Division.
\brlist
\bibliographystyle{unsrt}
\brf{LEP2} `Physics at LEP2,' edited by G. Altarelli, T. Sj\"ostrand
               and F. Zwirner,\\
              CERN96-01 (Vol. 1), Feb 1996.\\
          See Section 7 (heavy-quark physics), p. 330,
                      in the chapter on $\gamma\,\gamma$ physics.
\brf{VDM}J.J. Sakurai and D. Schildknecht,
             Phys. Lett. {\bf B40} (1979) 121;\\
                       I. F. Ginzburg and V.G. Serbo,
                            Phys. Lett. {\bf B109} (1982) 231.
\brf{CSM}R. M. Godbole {\it et al.}, Phys. Rev. {\bf D65} (2002) 074003;\\
   M. Klasen {\it et al.}, Nucl. Phys. {\bf B609} (2001) 518; \\
   B. Naroska, Nucl. Phys. B (Proc. Suppl.) {\bf 82} (2000) 187; \\
   H. Jung, G. A. Schuler and J. Terr\'on,
         Int. J. Mod. Phys. {\bf A7} (1992) 7955; \\
   E. L. Berger and D. Jones, Phys. Rev. {\bf D23} (1981) 1521.
         
\brf{DELPHI1}P. Aarnio {\it et al.},
             DELPHI Collab., Nucl. Inst. Meth. {\bf A303} (1991) 233.
\brf{DELPHI2}P. Abreu {\it et al.},
             DELPHI Collab., Nucl. Inst. Meth. {\bf A378} (1996) 57.
\brf{L3}R. Acciarri  {\it et al.}, L3 Collab., Phys. Lett. {\bf B503} (2001) 
10.   
\brf{PDG}    Particle Data Group,
             K.Hagiwara et al., Phys. Rev. {\bf D66}, 010001 (2002).
\brf{PYTHIA}T. Sj\"{o}strand {\it et al}, Comput. Phys. Comm. {\bf 135} (2001)
 238.
\erlist
\end{document}